\documentclass[prl,twocolumn,showpacs,superscriptaddress,preprintnumbers,amsmath,amssymb,floatfix]{revtex4}
\usepackage{times}
\usepackage{graphicx}% Include figure files
\usepackage{dcolumn}% Align table columns on decimal point
\usepackage{bm}% bold math
\usepackage{subfigure}

\begin{document}
\title{Barrier control in tunneling $\bm{e}^{\bm{+}}\text{-}\bm{e}^{\bm{-}}$ photoproduction}
\author{A. Di~Piazza}
\email{dipiazza@mpi-hd.mpg.de}
\affiliation{Max-Planck-Institut f\"{u}r Kernphysik, 
Postfach 103980, 69029 Heidelberg, Germany}
\author{E. L{\"o}tstedt}
\affiliation{Max-Planck-Institut f\"{u}r Kernphysik, 
Postfach 103980, 69029 Heidelberg, Germany}
\author{A. I.~Milstein}
\affiliation{Max-Planck-Institut f\"{u}r Kernphysik, 
Postfach 103980, 69029 Heidelberg, Germany}
\affiliation{Budker Institute of Nuclear Physics, 630090 Novosibirsk, Russia}
%\altaffiliation{Permanent address: Budker Institute of Nuclear Physics, 630090 Novosibirsk, Russia}
\author{C. H.~Keitel}
\affiliation{Max-Planck-Institut f\"{u}r Kernphysik, 
Postfach 103980, 69029 Heidelberg, Germany}
\date{\today}
{\begin{abstract}
Tunneling electron-positron pair production is studied in a new setup
in which a strong low-frequency and a weak high-frequency laser field 
propagate in the same direction and collide head-on with a relativistic nucleus.
The electron-positron pair production rate is calculated analytically in the 
limit in which in the nucleus rest frame
the strong field is undercritical and
the frequency of the weak field is below and 
close to the pair production threshold. By changing the frequency of the weak field one can reduce
the tunneling barrier substantially. As a result tunneling pair production is
shown to be observable with presently available technology.
\end{abstract}}

%%%%%%%%%%%%%%%%%%%%%%%%%%%%%%%%%%%%%%%%%%%%%%%%%%%%%%%%%%%%%%%%%%%%%%%
\pacs{
12.20.Ds, 	%Specific calculations in QED
%12.20.Fv, 	%Experimental tests of QED
25.75.Dw, 	%Particle and resonance production
%32.80.Wr, 	%Other multiphoton processes
%42.50.Xa, 	%Optical tests of quantum theory
42.62.-b 	%Laser applications
}
\maketitle
%%%%%%%%%%%%%%%%%%%%%%%%%%%%%%%%%%%%%%%%%%%%%%%%%%%%%%%%%%%%%%%%%%%%%%%%%
Electron-positron ($e^+\text{-}e^-$) pair creation from vacuum in the presence of
a constant and uniform electric field was predicted 
for the first time in the paper \cite{Sauter_1931a} 
(see also \cite{Heisenberg_1936,Schwinger_1951}). 
The typical electric field strength at which spontaneous
$e^+\text{-}e^-$ pair creation from vacuum occurs is now known as the ``critical'' 
field of quantum electrodynamics (QED) and it is given by $E_{cr}=m^2/e=1.3\times 10^{16}$ V/cm.
Here $e$ and $m$ are the absolute value of the electron charge and the electron mass, respectively, and
units with $\hbar=c=1$ are used. It is remarkable that the probability of pair creation contains
a non-perturbative dependence on the electric field amplitude $E$ and on the charge $e$ through the factor 
$\exp(-\pi E_{cr}/E)$. This suggests the interpretation of the process as a ``tunneling'' of the electron 
through an energy barrier of $2m$ from the negative energy levels of the Dirac ``sea'' to the 
positive ones \cite{Heisenberg_1936}. Moreover, this result cannot be
obtained at any order in perturbative QED and it can represent a truly deep
probe of the validity of QED. However, tunneling pair production has
not yet been observed experimentally essentially due to the wide tunneling barrier $2m$
and consequently to the large value of $E_{cr}$.

High-power lasers are a source of intense electromagnetic fields and nowadays peak
electric fields of the order of $10^{-4}E_{cr}$ have been obtained 
corresponding to laser intensities of the order of $10^{22}\;\text{W/cm$^2$}$ \cite{Emax}.
Moreover, Petawatt laser systems are under construction aiming at laser intensities
of the order of $10^{23}\;\text{W/cm$^2$}$ \cite{Norby_2005}. Finally,
intensities of the order of $10^{26}\;\text{W/cm$^2$}$ are envisaged at the Extreme
Light Infrastructure (ELI) \cite{ELI_Laser}. $e^+\text{-}e^-$ pair creation
in a single plane wave is forbidden by energy-momentum conservation \cite{Schwinger_1951},
however it has been investigated instead theoretically in the collision of a photon and a
plane wave \cite{Re1962, NiRi1964}, of a nucleus and a plane wave  \cite{Yakovlev_1966,MiMuHaJeKe2006,KuRo2007}
and also in the head-on collision of two equal laser beams \cite{BuNaMuPo2006,bell:200403,Rufetal2009} 
(see also the recent reviews \cite{Reviews} for further references).
In this last case, since the pair creation process is confined in a space
region of the order of a Compton wavelength $\lambda_c=1/m=3.8\times 10^{-11}\;\text{cm}$, the resulting 
standing wave originating by the superposition
of the two counterpropagating plane waves is often approximated as a time-dependent
electric field \cite{Brezin_1970,Di_Piazza_2004,Schuetzhold_2008,Hebenstreit_2009}. If $E$ is the peak
electric field, assumed to be much smaller than $E_{cr}$, and $\omega$ its carrier angular frequency, the parameter $\xi=eE/m\omega$
determines the regime of pair production \cite{Brezin_1970}. On the one hand, the parameter $\xi$ can be interpreted as the ratio of the external field oscillation period and the typical pair formation time. Therefore, if $\xi\gg 1$ the field is almost constant during the pair-production process and 
the production probability scales as in the constant-field case, i. e. as $\exp(-\pi E_{cr}/E)$ characteristic for the tunneling regime \cite{Schwinger_1951}. On the other
hand, the parameter $\xi$ can also be interpreted as the work carried out by the external electric field on the electron in one Compton wavelength
divided by the photon energy $\omega$. Therefore, in the opposite limit $\xi\ll 1$ photon exchanges with the external field are unlikely and 
the field itself can be treated perturbatively. The pair-production process occurs in this limit essentially with the absorption of $2m/\omega$ photons from the external field and the pair-production rate scales as $\xi^{4m/\omega}$ corresponding to the multiphoton regime \cite{Brezin_1970}.

The only currently feasible proposals to observe laser-induced pair creation in ion-laser collision have been in the multiphoton regime \cite{Mu2009},
while in the collision of two laser beams, intensities at least of the order of $10^{24}\;\text{W/cm$^2$}$ 
are required \cite{bell:200403}. Experimental evidence of $e^+\text{-}e^-$ pair creation has
been reported in \cite{Buetal1997} where this process was observed in the multiphoton regime and in \cite{Chen_2009} where the
large pair yield measured was predominately due to the Bethe-Heitler process. 

In this Letter we put forward a realistic scheme to observe tunneling $e^+\text{-}e^-$ pair creation in the head-on collision
of a relativistic nucleus with a strong, low-frequency and a weak, high-frequency laser field 
that propagate in the same direction (see Fig. 1a). 
\begin{figure}
\begin{center}
\includegraphics*[width=0.99\columnwidth]{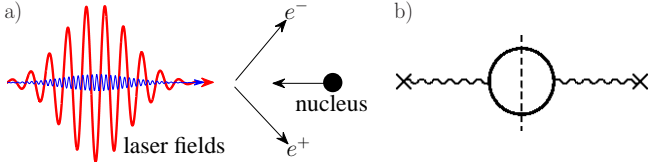}
\end{center}
\caption{\label{fig:schematic_setup}
(color online). Part a) Schematic setup of the considered process. An $e^+\text{-}e^-$ 
pair is created by a light pulse consisting of
a weak, high-frequency laser field and a strong, low-frequency 
laser field colliding head-on with a relativistic nucleus. Part b) Polarization operator of a photon in a plane wave. The crossed photon lines represent the Coulomb field of the nucleus and the thick electron lines are electron propagators in the plane wave. The vertical dashed line links the polarization operator to the pair production diagram.}
\end{figure}
The pair-creation rate is calculated analytically taking into account exactly the strong field and to leading order the weak and the nuclear field 
in the limit in which in the rest frame of the nucleus the peak electric field of the strong laser is much smaller than $E_{cr}$
and the frequency $\omega_{\text{w}}$ of the weak field is close to and below the pair creation threshold $2m$. In the limit of $\omega_{\text{w}}\gg m$ it is found that an external field suppresses the photoproduction yield because it substantially reduces the formation or coherence length of the process \cite{Baier_2005}. Instead, we find here that the strong laser field allows the process by making the electron tunnel the residual energy gap $2m-\omega_{\text{w}}$ left after the electron has absorbed one photon from the high-frequency field. By changing the frequency $\omega_{\text{w}}$ one can then control the amplitude of the barrier that the electron has to tunnel with the possibility of reducing it significantly and correspondingly enhance the pair production rate. Also, a strong dependence of the pair production rate on the mutual polarization of the two laser fields is observed as the absorption of one high-energy photon and the subsequent laser-induced tunneling effectively couple the polarization states of the two fields. Finally, a quantitative estimate suggests the possibility of observing tunneling pair production in the discussed regime with available laser and proton accelerator technology. 

The model case of pair creation in two parallel time-depending electric fields, one strong and slowly-varying and the other weak and rapidly-changing has been considered in \cite{Schuetzhold_2008}. However, the frequency of the weak field was assumed in the calculations to be much smaller than $2m$
leaving out the possibility of strongly reducing the energy barrier to be tunneled by the electron.

The total $e^+\text{-}e^-$ pair production rate $\dot{W}$ in the Born approximation with respect to the Coulomb field of the nucleus and being exact with regard to a general plane wave (to be chosen below as the sum of a strong, low-frequency wave and a weak, high-frequency wave) can be calculated by employing the dispersion relation, which allows one to express $\dot{W}$ via the imaginary part of the time-time component $\Pi^{00}$ of the polarization operator of a virtual photon in the plane wave field \cite{Landau_b_4} (see also Fig. 1b): 
\begin{equation}
\dot{W}=\frac{(4\pi Ze)^2}{4\pi}\int\frac{d^3q}{(2\pi)^3}\,\frac{\text{Im}\Pi^{00}}{|\mathbf{q}|^4}.
\end{equation}
Here $Z$ is the nuclear charge number and $\mathbf{q}$ denotes the momentum transferred from the nucleus during the process. Note that the Coulomb field can be accounted for in the Born approximation if $Z\alpha/u\ll 1$, where $\alpha=e^2\approx 1/137$ is the fine-structure constant and $u$ is the typical velocity of the created electron and positron. The calculations are performed in the rest frame of the nucleus. A convenient expression for $\Pi^{00}$ was obtained in \cite{BaMiSt1975} by means of the operator technique (the polarization operator in another form was obtained independently in \cite{BeMi1975}). In the general case the incoming electromagnetic field is described by a plane wave with vector potential $\mathbf{A}(\phi)=\mathbf{a}_1\psi_1(\phi)+\mathbf{a}_2\psi_2(\phi)$, where $\psi_i(\phi)$ with $i\in\{1,\,2\}$ are two arbitrary functions of $\phi=t-z$ (the plane wave depends on time $t$ and propagates in the positive $z$ direction with unity vector $\hat{\mathbf{z}}$), and $\mathbf{a}_i$ are the two polarization vectors such that $\mathbf{a}_i\cdot\mathbf{a}_j=0$ for $i,j\in\{1,\,2\}$ with $i\neq j$ and $\mathbf{a}_i\cdot\hat{\mathbf{z}}=0$. By employing the expression of $\Pi^{00}$ from \cite{BaMiSt1975}, we obtain
\begin{equation}
\label{W_dot_i}
\begin{split}
&\dot{W}=-\frac{(Z\alpha)^2m}{\pi^2}\text{Im}\int_0^{\infty}\frac{dQ}{Q^2}\int_0^1dv\int_0^{\infty}\frac{d\rho}{\rho}\int_0^T\frac{d\phi}{T}\int_0^1dx\\
&\times\frac{1-x^2}{x^2}
\exp\left\{-i\frac{a\rho}{Qx(1-v^2)}[1+Q^2(1-v^2)+\beta]\right\}\\
&\times\left[\frac{3-v^2}{1-v^2}\mathbf{\Delta}(1)\cdot\mathbf{\Gamma}-\mathbf{\Delta}^2(1)-F\right],
\end{split}
\end{equation}
where the integration $\int_0^T d\phi/T$ corresponds to the average of the integrand in the above equation over $\phi$ during the time period $T$ (later on set equal to the strong laser period), $Q=|\mathbf{q}|/2m$ and $x=\hat{\mathbf{z}}\cdot\mathbf{q}/|\mathbf{q}|$. Also, the following notation has been introduced: $F=1+Q^2[1-3v^2+x^2(1+v^2)]/(1-x^2)$, $\mathbf{\Delta}(y)=(e/m)[\mathbf{A}(\phi-\rho y/\omega_{\text{w}})-\mathbf{A}(\phi)]$, $\mathbf{\Gamma}=\int_0^1dy\mathbf{\Delta}(y)$, $\beta=\mathbf{\Gamma}^2-\int_0^1dy\mathbf{\Delta}^2(y)$ and $a=2m/\omega_{\text{w}}$. In the case of interest here, the plane electromagnetic field $\mathbf{A}(\phi)$ consists of a strong, monochromatic, low-frequency field with adimensional vector potential components in the plane perpendicular to $\hat{\mathbf{z}}$ $\xi_i=eE_i/m\omega_s$ with $i\in\{1,\,2\}$ and a weak, monochromatic, high-frequency field with adimensional vector potential components $\eta_i=e\mathcal{E}_i/m\omega_{\text{w}}$. $E_i$ and $\mathcal{E}_i$ are the electric field components of the strong and the weak field, respectively and $\omega_s$ is the strong field angular frequency. Concerning the strong field it is assumed that $\xi=\sqrt{(\xi_1^2+\xi_2^2)/2}\gg 1$ and $\omega_s\ll m$, while concerning the weak field that $\eta=\sqrt{(\eta_1^2+\eta_2^2)/2}\ll 1$ and that $0<\delta\ll 1$ with $\delta= a^2-1\approx (2m-\omega_{\text{w}})/m$. These assumptions are reasonable for already and soon available laser sources in the optical and X-ray regime. In the above approximations for the strong field the rate $\dot{W}$ depends on the strong field only through the two gauge- and Lorentz-invariant parameters
\begin{equation}
\chi_i= \frac{\omega_s}{\omega_{\text{w}}} \xi_i=\frac{E_i}{E_{cr}}\frac{m}{\omega_{\text{w}}}.
\end{equation}
It is convenient to introduce the quantities
\begin{equation}
\chi=\sqrt{\frac{\chi_1^2+\chi_2^2}{2}}, \quad \mu_s=\frac{\chi_1^2-\chi_2^2}{\chi_1^2+\chi_2^2}, \quad \mu_{\text{w}}=\frac{\eta_1^2-\eta_2^2}{\eta_1^2+\eta_2^2},
\end{equation}
with the parameters $\mu_s$ and $\mu_{\text{w}}$ describing the ellipticities of the strong and of the weak field, respectively.

As $\eta\ll 1$, the weak laser field can be treated perturbatively and the general expression in Eq. (\ref{W_dot_i}) for the photoproduction rate $\dot{W}$ can be expanded with respect to the amplitudes $\eta_i$ by keeping only the terms quadratic in $\eta_i$, which correspond to the absorption of one photon from the weak field. The result is:
\begin{widetext}
\begin{equation}
\label{W_dot_t}
\begin{split}
\dot{W}&=-\frac{(Z\alpha)^2m\eta^2}{\pi^2}\text{Im}\int_0^{\infty}\frac{dQ}{Q^2}\int_0^1dv\int_0^{\infty}\frac{d\rho}{\rho}\int_0^{2\pi}\frac{d\psi}{2\pi}\int_0^1dx\frac{1-x^2}{x^2}e^{-i\Phi}\left\{i\frac{a\rho}{Qx(1-v^2)}\left[1-\frac{\sin^2\rho}{\rho^2}-i\frac{a\rho\chi^2}{Qx(1-v^2)}gR^2\right]\right.\\
&\times
\left.\left(F-2\rho^2\chi^2f\frac{1+v^2}{1-v^2}\right)-\frac{1+v^2}{1-v^2}\sin\rho\left[i\frac{4a\rho^2\chi^2}{Qx(1-v^2)}gR-2\sin\rho\right]\right\},
\end{split}
\end{equation}
\end{widetext}
where $\Phi=a\rho[1+f\chi^2\rho^2/3+Q^2(1-v^2)]/Qx(1-v^2)$, $R=\sin\rho\,/\rho-\cos\rho$, $f=1-\mu_s\cos\psi$ and $g=1+\mu_{\text{w}}\mu_s-(\mu_s+\mu_{\text{w}})\cos\psi$. Note that, as expected, the rate depends linearly on the parameter $\mu_{\text{w}}$ describing the ellipticity of the weak field while the dependence on $\mu_s$ is more complex. In the limit in which the strong field is undercritical ($\chi\ll 1$) and the frequency of the weak field is below and close to $2m$ ($0<\delta\ll 1$) the integrals in Eq. (\ref{W_dot_t}) can be calculated by employing the saddle-point technique. The resulting rate $\dot{W}$ reads:
\begin{equation}
\label{W_dot_tf}
\dot{W}=\frac{(Z\alpha)^2m\eta^2\chi^2}{16\sqrt{\pi}}\sqrt{\zeta}\int_0^{2\pi}\frac{d\psi}{2\pi}\mathcal{G}\exp\left(-\frac{2}{3\zeta\sqrt{f}}\right),
\end{equation}
where $\mathcal{G}=f^{1/4}\left(g+2\zeta f^{3/2}\right)$ and where we have introduced the important parameter $\zeta=\chi/\delta^{3/2}$. The expression in Eq. (\ref{W_dot_tf}) is valid if $\zeta\ll 1$. Formally $\zeta f^{3/2}\ll g$, however, if the strong laser field and the high-energy photon are linearly polarized in perpendicular directions, we have $g=0$ and the non-zero contribution to $\dot{W}$ is given by the term $2\zeta f^{3/2}$. It is already clear from Eq. (\ref{W_dot_tf}) that the exponential suppression associated with the tunneling nature of pair creation is strongly reduced here because $\delta\ll 1$. In the case of combined strong low-frequency laser and Coulomb field the pair production rate scales as $\exp(-\sqrt{3}/\chi)$ \cite{MiMuHaJeKe2006}. Also, contrary to the rate in \cite{MiMuHaJeKe2006}, our expression (\ref{W_dot_tf}) also contains the small factor $\eta^2$ due to the absorption of one photon from the weak field. Therefore, the condition to be fulfilled in order to have an enhancement of the pair production rate with respect to the case of combined laser and Coulomb field is approximately given by $\eta^2\exp(\sqrt{3}/\chi)\gg 1$. Analogously, in \cite{Kalman} atomic ionization in the presence of a strong, optical field and of a weak, X-ray field with photon-energy below and close to the ionization threshold has been investigated.

The above expression (\ref{W_dot_tf}) is valid for any polarization of the incoming strong and weak fields. In particular, if the strong laser field is circularly polarized ($\mu_s=0$), we obtain:
\begin{equation}
\label{sigma_c}
\dot{W}=\frac{(Z\alpha)^2m\eta^2\chi^2}{16\sqrt{\pi}}\sqrt{\zeta}\exp\left(-\frac{2}{3\zeta}\right).
\end{equation}
As expected from symmetry considerations, this expression is independent of the polarization of the weak field. On the other hand, if $\mu_s$ is not too small, i. e. if $|\mu_s|\gg \zeta$ we have
\begin{equation}
\label{sigma_l}
\begin{split}
\dot{W}&=\frac{\sqrt{3}(Z\alpha)^2m \eta^2\chi^2}{16\pi\sqrt{2}}\zeta\frac{\kappa^2}{\sqrt{\kappa-1}}L\exp\bigg(-\frac{2}{3\zeta\sqrt{\kappa}}\bigg),
\end{split}
\end{equation}
with $L=1+\text{sgn}(\mu_s)\mu_{\text{w}}+2\zeta\sqrt{\kappa}$ and $\kappa=1+|\mu_s|$. 

\begin{figure}
\begin{center}
\includegraphics*[width=0.6\columnwidth]{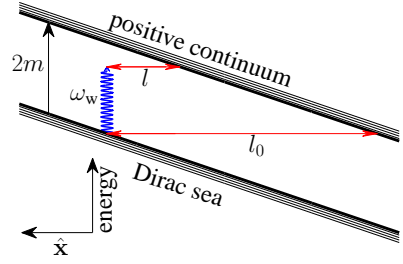}
\end{center}
\caption{\label{fig:tunnel}
(color online). Schematic picture of the tunneling mechanism. The electron has initially
a negative energy $-m$, absorbs one photon with energy $\omega_{\text{w}}$, and is finally transferred into the positive continuum by tunneling the
distance $l=l_0-\omega_{\text{w}}/eE_0$ with $l_0=2m/eE_0$ through the barrier tilted by the electric field $\mathbf{E}_0=E_0\hat{\mathbf{x}}$.}
\end{figure}

For an intuitive digression we show qualitatively how the exponential behavior in Eqs. (\ref{sigma_c}) and (\ref{sigma_l}) arises (see also \cite{Schuetzhold_2008}). We consider the simplified situation of an electron in the negative continuum that absorbs a photon with frequency $\omega_{\text{w}}$ and then, due to a constant and uniform electric field $\mathbf{E}_0=E_0\hat{\mathbf{x}}$ with $E_0>0$, it tunnels to the positive continuum, as depicted in Fig. \ref{fig:tunnel}. The width $l$ of the barrier the electron has to tunnel is approximately given by the formula $eE_0l=2m-\omega_{\text{w}}\approx m\delta\ll m$ (note that $l\ll l_0$, with $l_0=2m/eE_0$ being the tunneling width only in the presence of the field $\mathbf{E}_0$). Therefore, by setting $p(x)=\sqrt{2m(m\delta-eE_0x)}$, in the quasiclassical limit one obtains
\begin{equation}
\frac{\dot{W}}{m}\sim\exp\bigg[-2\int_0^ldx\, p(x)\bigg]=\exp\bigg(-\frac{2\sqrt{2}}{3\zeta_0}\bigg),
\end{equation}
with $\zeta_0=E_0/2\delta^{3/2}E_{cr}$ being defined analogously to $\zeta$ and $\zeta_0\ll 1$, which qualitatively reproduces the exponential dependence in Eqs. (\ref{sigma_c}) and (\ref{sigma_l}). Also, from the above picture we deduce that the velocity $u$ of the electron before tunneling is nonrelativistic as $u=\sqrt{2(2m-\omega_{\text{w}})/m}=\sqrt{2\delta}\ll 1$ and the formation time $\Delta t$ of the process is approximately $\Delta t=l/u\sim(\sqrt{\delta}/m)(E_{cr}/E_0)$, as from the uncertainty principle the time to absorb the photon with energy $\omega_{\text{w}}$ is roughly $1/\omega_{\text{w}}\sim 1/m\ll \Delta t$.

In Fig. 3 we display the ratio $\dot{W}/\dot{W}_0$ with $\dot{W}_0=(Z\alpha)^2m\eta^2\chi^2/2\pi$ and $\dot{W}$ given by Eq. (\ref{sigma_c}) for circular polarization of the strong laser field (continuous curve) and by Eq. (\ref{sigma_l}) for linear polarization of the strong and the weak laser field with $\mu_s=\mu_{\text{w}}=+1$ (dashed curve). 
\begin{figure}
%\begin{center}
\includegraphics[width=0.8\columnwidth]{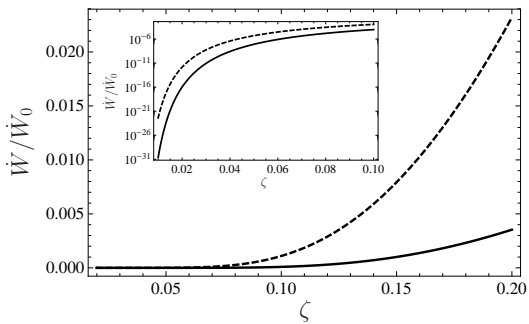}
%\end{center}
\caption{The ratio $\dot{W}/\dot{W}_0$ with $\dot{W}_0=(Z\alpha)^2m\eta^2\chi^2/2\pi$ and $\dot{W}$ given by Eq. (\ref{sigma_c}) for circular polarization of the strong laser field (continuous curve) and by Eq. (\ref{sigma_l}) for linear polarization of the strong and the weak laser fields with $\mu_s=\mu_{\text{w}}=+1$ (dashed curve). The inset shows the region of small $\zeta$ on a logarithmic scale.}
\end{figure}
It becomes apparent that the pair production rate is much higher in the case of linear polarization. This can also be understood from the different exponential scaling in Eqs. (\ref{sigma_c}) and (\ref{sigma_l}) which, in turn, is due the fact that at a given $\chi$ the average energy density of the strong laser is fixed and the peak electric field amplitude for linear polarization is $\sqrt{2}$ times larger than for circular polarization.

As an example, we consider a proton ($Z=1$) with an energy of $\epsilon_p=2.8\;\text{TeV}$ which is smaller than those maximally available at the LHC \cite{PDG} and a laser field with a power of $P_s=100\;\text{TW}$, a spot radius of $\sigma_s=5\;\text{$\mu$m}$ (intensity $I_s=1.3\times 10^{20}\;\text{W/cm$^2$}$), a wavelength of $\lambda_s=0.8\;\text{$\mu$m}$, a pulse duration of $\tau_s=25\;\text{fs}$ and a repetition rate of $f_r=10\;\text{Hz}$ corresponding to $\chi=7.5\times 10^{-2}$ \cite{Pittman_2002}. If we set $\delta=0.1$ then $\omega_{\text{w}}=162\;\text{eV}$ (note that the Born approximation in the Coulomb field is justified as $Z\alpha/\sqrt{2\delta}\approx 0.02$). As for the weak field we consider the following parameters \cite{FLASH}: $N_{\text{w}}=10^{13}$ photons per pulse, a pulse duration of $\tau_{\text{w}}=25\;\text{fs}$ and a spot radius of $\sigma_{\text{w}}=5\;\text{$\mu$m}$ and, in the most favorable case $\mu_s=\mu_{\text{w}}=+1$, we obtain a rate in the laboratory frame of $\dot{W}=750\;\text{s$^{-1}$}$. Since $\zeta=2.4$, the analytical asymptotic in Eq. (\ref{sigma_l}) slightly overestimates the rate and the above value of $\dot{W}$ has been obtained numerically directly from Eq. (\ref{W_dot_t}). By considering the values of the proton beams at the LHC ($N_p=11.5\times 10^{10}$ protons per bunch, bunch length of $l_p=7.55\;\text{cm}$ and beam size of $\sigma_p=16.6\;\text{$\mu$m}$ \cite{PDG}), we obtain about 18 pairs per hour. Note that in the case of combined Coulomb and strong laser field alone, the number of produced pairs is completely negligible with the above physical parameters (a larger rate than here can be expected at $\omega_{\text{w}}\gtrsim 2m$, but in this regime no tunneling occurs as pair creation is also allowed in the absence of the strong field). With the maximal proton energy available at the LHC of $\epsilon_p=7\;\text{TeV}$ and at $\delta=0.1$ then $\omega_{\text{w}}=65\;\text{eV}$. Thus, the intense single extreme ultraviolet (XUV) attosecond pulse ($N_{\text{w}}=10^{16}$, $\omega_{\text{w}}=65\;\text{eV}$, $\tau_{\text{w}}=84\;\text{as}$, $\sigma_{\text{w}}=10\;\text{$\mu$m}$) envisaged in \cite{Tsakiris_2006} can be employed as a more compact source of high-energy photons. This XUV pulse is generated in the reflection of a strong optical laser beam ($\lambda_s=0.8\;\text{$\mu$m}$, $I_s=10^{20}\;\text{W/cm$^2$}$, $\tau_s=5\;\text{fs}$, $\sigma_s=10\;\text{$\mu$m}$) by a planar solid target. By employing these parameters we obtain even a yield of about 13 pairs per shot. Finally, about one pair every ten hours can also be obtained at $\delta=0.1$ by combining the already operative accelerator Tevatron ($\epsilon_p=980\;\text{GeV}$, $N_p=2.4\times 10^{11}$, $l_p=50\;\text{cm}$, $\sigma_p=29\;\text{$\mu$m}$), a Petawatt laser ($I_s=10^{21}\;\text{W/cm$^2$}$, $\lambda_s=1.2\;\text{$\mu$m}$, $\sigma_s=5\;\text{$\mu$m}$, $\tau_s=4\;\text{fs}$, $f_r=10\;\text{Hz}$ \cite{PFS}) and the table-top X-FEL envisaged by the experimentalists in \cite{Gruener_2007} ($N_{\text{w}}=8\times 10^{11}$, $\omega_{\text{w}}=470\;\text{eV}$, $\tau_{\text{w}}=4\;\text{fs}$) with spot radius of $\sigma_{\text{w}}=5\;\text{$\mu$m}$.

In conclusion, we have put forward a scheme allowing in principle to observe for the first time tunneling electron-positron pair creation with already available technology, by inducing a strong, low-frequency and a weak, high-frequency laser field to collide with a relativistic nucleus.

The work was supported in part by the RFBR under the grants 09-02-00024 and 08-02-91969.

%\bibliography{/home/theo/lotstedt/samba/user/bibfiles/aaadefs,/home/theo/lotstedt/samba/user/bibfiles/eriksbib_2,/home/theo/lotstedt/samba/user/bibfiles/my_papers}

\begin{thebibliography}{39}
\expandafter\ifx\csname natexlab\endcsname\relax\def\natexlab#1{#1}\fi
\expandafter\ifx\csname bibnamefont\endcsname\relax
  \def\bibnamefont#1{#1}\fi
\expandafter\ifx\csname bibfnamefont\endcsname\relax
  \def\bibfnamefont#1{#1}\fi
\expandafter\ifx\csname citenamefont\endcsname\relax
  \def\citenamefont#1{#1}\fi
\expandafter\ifx\csname url\endcsname\relax
  \def\url#1{\texttt{#1}}\fi
\expandafter\ifx\csname urlprefix\endcsname\relax\def\urlprefix{URL }\fi
\providecommand{\bibinfo}[2]{#2}
\providecommand{\eprint}[2][]{\url{#2}}

\bibitem{Sauter_1931a} F. Sauter, Z. Phys. \textbf{69}, 742 (1931).
\bibitem{Heisenberg_1936} W. Heisenberg and H. Euler, Z. Phys. \textbf{98}, 714 (1936).
\bibitem{Schwinger_1951} J. Schwinger, Phys. Rev. \textbf{82}, 664 (1951).


\bibitem{Emax} V. Yanovsky et al., Opt. Express {\bf 16}, 2109 (2008).
\bibitem{Norby_2005} J. Norby, Laser Focus World, January 1 (2005).
\bibitem{ELI_Laser} http://www.extreme-light-infrastructure.eu/.

\bibitem[{\citenamefont{Reiss}(1962)}]{Re1962}
\bibinfo{author}{\bibfnamefont{H.~R.} \bibnamefont{Reiss}},
  \bibinfo{journal}{J. Math. Phys.} \textbf{\bibinfo{volume}{3}},
  \bibinfo{pages}{59} (\bibinfo{year}{1962}).

\bibitem[{\citenamefont{Nikishov and Ritus}(1964)}]{NiRi1964}
\bibinfo{author}{\bibfnamefont{A.~I.} \bibnamefont{Nikishov}} \bibnamefont{and}
  \bibinfo{author}{\bibfnamefont{V.~I.} \bibnamefont{Ritus}},
  \bibinfo{note}{Sov. Phys. JETP \textbf{19}, 529 (1964)}.

 \bibitem{Yakovlev_1966} V. P. Yakovlev, Sov. Phys. JETP \textbf{22}, 223 (1966).
\bibitem[{\citenamefont{Milstein et~al.}(2006)\citenamefont{Milstein, M\"uller,
  Hatsagortsyan, Jentschura, and Keitel}}]{MiMuHaJeKe2006}
\bibinfo{author}{\bibfnamefont{A.~I.} \bibnamefont{Milstein}} \bibnamefont{et~al.}, \bibinfo{journal}{Phys. Rev. A}
  \textbf{\bibinfo{volume}{73}}, \bibinfo{pages}{062106}
  (\bibinfo{year}{2006}).

\bibitem[{\citenamefont{Kuchiev and Robinson}(2007)}]{KuRo2007}
\bibinfo{author}{\bibfnamefont{M.~{\relax Yu}.} \bibnamefont{Kuchiev}}
  \bibnamefont{and} \bibinfo{author}{\bibfnamefont{D.~J.}
  \bibnamefont{Robinson}}, \bibinfo{journal}{Phys. Rev. A}
  \textbf{\bibinfo{volume}{76}}, \bibinfo{pages}{012107}
  (\bibinfo{year}{2007}).


\bibitem[{\citenamefont{Bulanov et~al.}(2006)\citenamefont{Bulanov, Narozhny,
  Mur, and Popov}}]{BuNaMuPo2006}
\bibinfo{author}{\bibfnamefont{S.~S.} \bibnamefont{Bulanov}} \bibnamefont{et~al.},
\bibinfo{note}{JETP  \textbf{102}, 9 (2006)}.

\bibitem[{\citenamefont{Bell and Kirk}(2008)}]{bell:200403}
\bibinfo{author}{\bibfnamefont{A.~R.} \bibnamefont{Bell}} \bibnamefont{and}
  \bibinfo{author}{\bibfnamefont{J.~G.} \bibnamefont{Kirk}},
  \bibinfo{journal}{Phys. Rev. Lett.} \textbf{\bibinfo{volume}{101}},
  \bibinfo{eid}{200403} (\bibinfo{year}{2008}).

\bibitem[{\citenamefont{Ruf et~al.}(2009)\citenamefont{Ruf, Mocken, M\"{u}ller,
  Hatsagortsyan, and Keitel}}]{Rufetal2009}
\bibinfo{author}{\bibfnamefont{M.}~\bibnamefont{Ruf}} \bibnamefont{et~al.}, \bibinfo{journal}{Phys. Rev. Lett.}
  \textbf{\bibinfo{volume}{102}}, \bibinfo{eid}{080402}
   (\bibinfo{year}{2009}).
\bibitem{Reviews} G. A. Mourou, T. Tajima, and S. V. Bulanov, Rev. Mod. Phys. \textbf{78}, 309 (2006); 
M. Marklund and P. K. Shukla, \textit{ibid.}, 591 (2006); Y. I. Salamin et al., Phys. Rep. \textbf{427}, 41 (2006).


\bibitem{Brezin_1970} E. Br\'{e}zin and C. Itzykson, Phys. Rev. D \textbf{2}, 1191 (1970).
\bibitem{Di_Piazza_2004} A. Di Piazza, Phys. Rev. D \textbf{70}, 053013 (2004).
\bibitem{Hebenstreit_2009}  F. Hebenstreit et al., Phys. Rev. Lett. \textbf{102}, 150404 (2009).
\bibitem{Schuetzhold_2008} R. Sch\"{u}tzhold, H. Gies, and G. Dunne, Phys. Rev. Lett. \textbf{101}, 130404 (2008).

\bibitem[{\citenamefont{Muller}(2009)\citenamefont{C. M\"{u}ller}}]{Mu2009}
\bibinfo{author}{\bibfnamefont{C.}~\bibnamefont{M\"{u}ller}}, \bibinfo{journal}{Phys. Lett. B}
  \textbf{\bibinfo{volume}{672}}, \bibinfo{pages}{56}
  (\bibinfo{year}{2009}).


\bibitem[{\citenamefont{Burke et~al.}(1997)\citenamefont{Burke, Field,
  Horton-Smith, Spencer, Walz, Berridge, Bugg, Shmakov, Weidemann, Bula
  et~al.}}]{Buetal1997}
\bibinfo{author}{\bibfnamefont{D.~L.} \bibnamefont{Burke}} \bibnamefont{et~al.},
  \bibinfo{journal}{Phys. Rev. Lett.} \textbf{\bibinfo{volume}{79}},
  \bibinfo{pages}{1626} (\bibinfo{year}{1997}).

\bibitem{Chen_2009} H. Chen et al., Phys. Rev. Lett. \textbf{102}, 105001 (2009).

\bibitem{Baier_2005} V. N. Baier and V. M. Katkov, Phys. Rep. \textbf{409}, 261 (2005).

\bibitem{Landau_b_4} V. B. Berstetskii, E. M. Lifshitz, and L. P. Pitaevskii, \textit{Quantum Electrodynamics}
(Elsevier, Oxford, 1982), Sec. 71.

\bibitem[{\citenamefont{Ba{\u \i}er et~al.}(1975)\citenamefont{Ba{\u \i}er,
  Mil'ste{\u \i}n, and Strakhovenko}}]{BaMiSt1975}
\bibinfo{author}{\bibfnamefont{V.~N.} \bibnamefont{Ba{\u \i}er}},
  \bibinfo{author}{\bibfnamefont{A.~I.} \bibnamefont{Mil'ste{\u \i}n}},
  \bibnamefont{and} \bibinfo{author}{\bibfnamefont{V.~M.}
  \bibnamefont{Strakhovenko}}, 
  \bibinfo{note}{Sov. Phys. JETP {\bf 42}, 961 (1976)}.

\bibitem[{\citenamefont{Becker and Mitter}(1975)}]{BeMi1975}
\bibinfo{author}{\bibfnamefont{W.}~\bibnamefont{Becker}} \bibnamefont{and}
  \bibinfo{author}{\bibfnamefont{H.}~\bibnamefont{Mitter}},
  \bibinfo{journal}{J. Phys. A} \textbf{\bibinfo{volume}{8}},
  \bibinfo{pages}{1638} (\bibinfo{year}{1975}).

\bibitem{Kalman} P. K\'{a}lm\'{a}n, Phys. Rev. A \textbf{38}, 5458 (1988); \textit{ibid.}, \textbf{39}, 3200 (1989).

\bibitem{PDG} Particle Data Group: W.-M. Yao et al., J. Phys. G \textbf{33}, 1 (2006).
\bibitem{Pittman_2002} M. Pittman et al., Appl. Phys. B \textbf{74}, 529 (2002).

\bibitem{FLASH} http://hasylab.desy.de/facilities/flash/machine/index\_eng.html.
\bibitem{Tsakiris_2006} G. D. Tsakiris et al., New J. Phys. \textbf{8}, 19 (2006).


\bibitem{PFS} http://www.attoworld.de/research/PFS.html.
\bibitem{Gruener_2007} F. Gr\"uner et al., Appl. Phys. B \textbf{86}, 431 (2007).

\end{thebibliography}

\end{document}